# Laboratory unravelling of matter accretion in young stars

G. Revet[1,2], S.N. Chen[1,2], R. Bonito[3,4], B. Khiar[5], E. Filippov[6,7], C. Argiroffi[4,3], D. P. Higginson[2], S. Orlando[3], J. Béard[8], M. Blecher[9], M. Borghesi[10], K. Burdonov[1], D. Khaghani[11], K. Naughton[10], H. Pépin[12], O. Portugall[8], R. Riquier[2,13], R. Rodriguez[14], S.N. Ryazantsev[6,7], I.Yu. Skobelev[6,7], A. Soloviev[1], O. Willi[9], S. Pikuz[6,7], A. Ciardi[5], J. Fuchs[1,2,]*

**Affiliations:**
[1]Institute of Applied Physics, 46 Ulyanov Street, 603950 Nizhny Novgorod, Russia
[2]LULI - CNRS, École Polytechnique, CEA: Université Paris-Saclay; UPMC Univ Paris 06: Sorbonne Universités - F-91128 Palaiseau cedex, France
[3]INAF - Osservatorio Astronomico di Palermo, Italy
[4]Dipartimento di Fisica e Chimica Università di Palermo, Italy
[5]LERMA, Observatoire de Paris, CNRS UMR 8112, Paris France.
[6]National Research Nuclear University 'MEPhI', 115409 Moscow, Russia
[7]Joint Institute for High Temperatures, RAS, 125412, Moscow, Russia
[8]LNCMI, UPR 3228, CNRS-UFJ-UPS-INSA, 31400 Toulouse, France.
[9]Institut für Laser- und Plasmaphysik, Heinrich-Heine-Universität Düsseldorf, D-40225 Düsseldorf, Germany
[10] Centre for Plasma Physics, The Queen's University of Belfast, Belfast BT7 1NN, United Kingdom
[11]GSI Helmholtzzentrum für Schwerionenforschung GmbH, 64291 Darmstadt, Germany
[12]INRS-EMT, Varennes, Québec, Canada.
[13]CEA, DAM, DIF, 91297 Arpajon, France
[14]Departamento de Fisica de la Universidad de Las Palmas de Gran Canaria, E-35017 Las Palmas de Gran Canaria, Spain

*Correspondence to: julien.fuchs@polytechnique.fr.

**Abstract:**

Accretion dynamics in the forming of young stars is still object of debate because of limitations in observations and modelling. Through scaled laboratory experiments of collimated plasma accretion onto a solid in the presence of a magnetic field, we open first window on this phenomenon by tracking, with spatial and temporal resolution, the dynamics of the system and simultaneously measuring multiband emissions. We observe in these experiments that matter, upon impact, is laterally ejected from the solid surface, then refocused by the magnetic field toward the incoming stream. Such ejected matter forms a plasma shell that envelops the shocked core, reducing escaped X-ray emission. This demonstrates one possible structure reconciling current discrepancies between mass accretion rates derived from X-ray and optical observations.

**Main Text:**

*Introduction*

The accretion of matter is a process that plays a central role in varied astrophysical systems across the mass spectrum (*1,2*). It determines the exchange of mass, energy, and angular momentum between the accreting object and its surroundings, and, eventually, formation of planetary systems around stars. In the context of young stars, during the early stages of formation and evolution toward the main sequence phase, the final mass of the forming star is therefore determined by the accretion process. Hence, the analysis of accreting young stars and the structure by which matter settles on the star is of wide interest in the context of star and planet formation in general.

In our present understanding, referred to as the magnetospheric accretion model, matter (2000 K in temperature and $10^{11}$-$10^{13}$ cm$^{-3}$ electron plasma density) is accreted onto young stars through magnetized (B = 0.01 – 0.1 T) accretion columns that connect the surrounding material (from the envelope in the early phases or the edge of the disk in the classical T Tauri stage (*3*)) to the star's surface (*4,5*). Guided along the lines of the magnetic field that connects the surrounding material to the star, matter falls at free-fall velocity, i.e. with a speed of the order of 500 km/s, and impacts onto the stellar chromosphere. A shock forms at the interaction between the accretion stream and the chromosphere, and the post shock plasma is heated to a few millions degree (e.g., *6*), with a pre-shock velocity of the order of 500 km/s and a post-shock reduced speed of about 100 km/s. Therefore, an excess of emission in UV and X-rays is expected to originate from this hot plasma. Investigating the shock structure and the high energy emission from the central star and its components (corona, accretion shocks, and jets) is crucial as this emission can influence several aspects of the star-disk system: it can alter the physical and chemical structure of the disk around the young star, with effects also on the disk's lifetime; and it can inhibit the formation of exo-planetary systems; additionally, it by itself it can be used as a diagnostic tool to study the properties of the accreting material and the shock physics. Hence, analysing the emission from these stars, in particular in the X-ray band, has been thought to allow insight into the accretion phenomenon (*7,8*), as well as into mass accretion rates (e.g., *9,10,11*).

Mass accretion rates are inferred in several ways. This can be done, for instance, through the analysis of optical observations (e.g. *12*), i.e. by deducing the star accretion luminosity from its optical-emission line's luminosities through an empirical relation, after which the mass-accretion rate can be deduced from the accretion luminosity (*13*). More recently, the mass-accretion rate of Classical T Tauri stars (CTTSs) has been derived from their soft X-ray emission (e.g. *6,14,15*). For all methods, it is assumed that: 1) the stream impacts the stellar surface with free fall velocity from a distance equal to a few (typically five) times the radius of the star, and 2) the impact region is in stationary conditions, namely the velocity, density, and temperature of the shock-heated plasma do not change in time.

*Present issues with inferring mass accretion rates from observations of your stars*

However, observations from CTTSs, where the accretion process is significant, show that there are significant discrepancies (e.g., *11,16*) between the observed X-ray luminosity and predictions based on the UV/visible bands from the same object. In some systems, the discrepancy is moderate (i.e. below a factor of 4; e.g. Hen 3-600, TW Hya), while in others the discrepancy is large (up to two orders of magnitude; e.g. RU Lup, T Tau), but in all cases the observed X-ray luminosity is lower than expected. Based on magnetohydrodynamic (MHD) models (e.g., *24,5,23,17,18,19,20*), so far limited to two-spatial dimensions, absorption of X-rays by an optically thick envelope of plasma surrounding the shocked plasma (*5,21,22,23*) is one of several scenarios (*24*) that have been evoked to explain this discrepancy and the lower than expected observed X-ray luminosity. However direct observation of such an envelope is well beyond present-day observation capabilities.

*Laboratory experiments to resolve the accretion dynamics in scaled conditions*

An alternative approach to the investigation of astrophysically-relevant plasma dynamics is through scaled laboratory experiments (*25,26,27*). In the frame of accretion dynamics, as mentioned above, this requires mimicking in the laboratory accretion streams embedded in magnetic fields. Although laser-produced high-strength magnetic fields have been developed for some times (*28,29*), coupling quasi-static, externally applied (i.e. not self-generated) magnetic fields with laser-produced plasmas has only recently become possible (*30,31,32,33*). In order to model in the laboratory the accretion columns guided along the lines of the magnetic field that connects the surrounding material to the star, we exploit here an experimental platform we have recently developed (*25,34*) that ensures the to homogeneity of the externally-applied magnetic field over a very large volume (cubic cms). As will be shown below, this is a key factor in allowing the present observations as it ensures that a homogeneous magnetic field (and also of high-strength) exists over the scale of the accretion column impact and lateral expansion on the mimicked star surface.

Our experimental results obtained in this frame and with this setup (Fig 1C and Fig 2) provide the first direct evidence for the formation of a shell of dense ionized plasma enveloping the core post-shock region, which is well consistent with 3D MHD simulations of the laboratory experiment (Fig 1B) and with 2D MHD scaled astrophysical simulations (Fig 1D). Furthermore, when post-processing the astrophysical simulation (Fig 3), this shell is observed to induce absorption of X-rays arising from the ionized central core, thus supporting the scenario of X-ray flux lowering by local medium surrounding the shocked region (*22*).

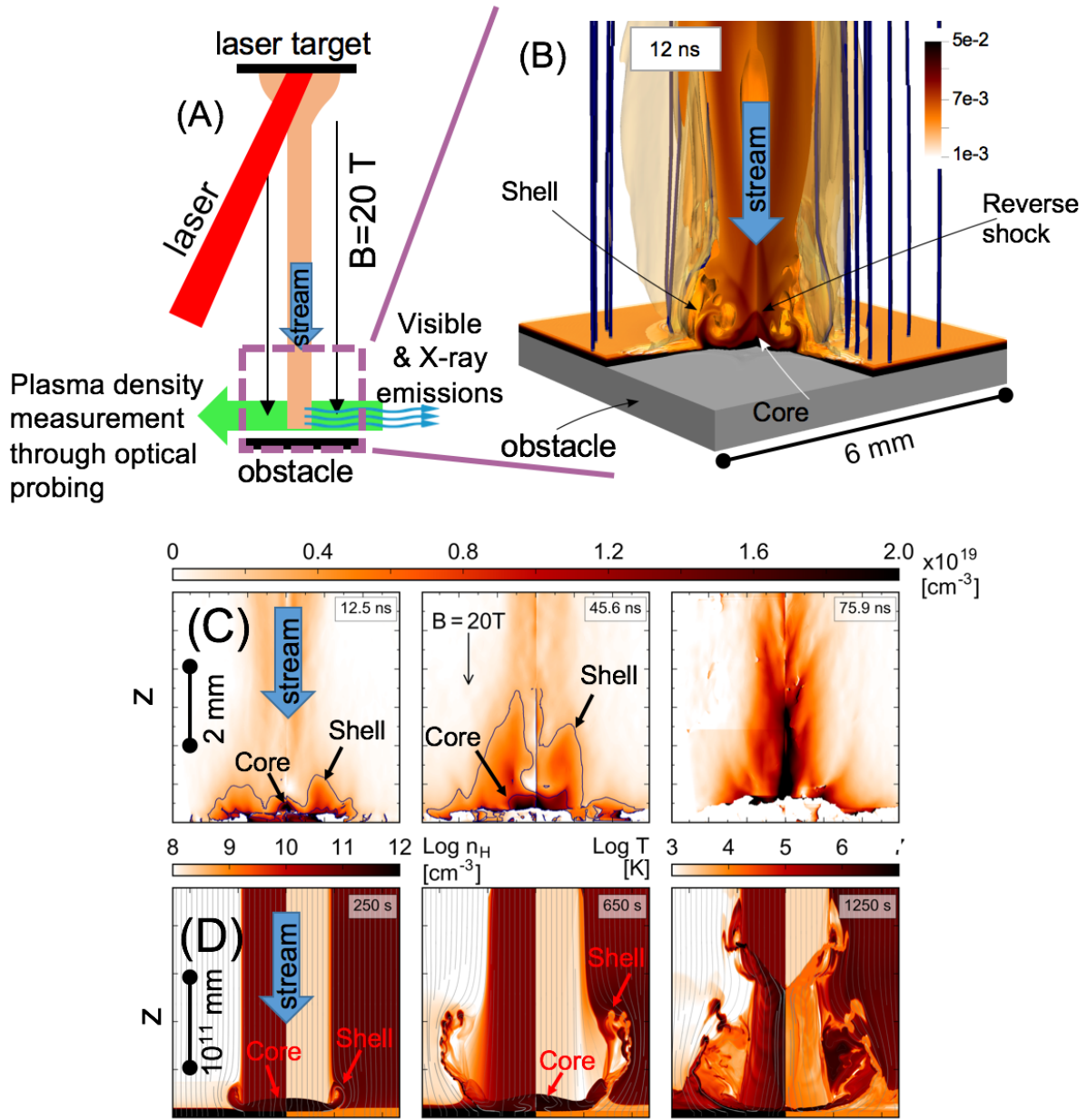

**Fig. 1: Laboratory investigation of magnetized accretion dynamics and comparison with scaled astrophysical simulation of the same phenomenon highlighting the formation of a shocked core and of a surrounding shell.** (A) Arrangement of the laboratory experiment and of the diagnostics. (B) Snapshot of the modelling of the laboratory experiment by the GORGON code (shown is the mass density in kg/m³). (C) Measured maps, at different times (as indicated), of laboratory plasma electron density, embedded in a homogeneous and steady 20 T magnetic field. The contours displayed on the two first panels highlight the core (contour at $10^{19} cm^{-3}$) and the

shell (contour at $3\times10^{18} cm^{-3}$). Here the obstacle is a $CF_2$ target, while the stream is generated from a laser-irradiated PVC $(C_2H_3Cl)_n$) target. (D) Simulated plasma density (left half-panels) and temperature (right half-panels) maps, also at different times, and extracted from a two-dimensional astrophysical simulation (using the PLUTO code). In all panels, the initial magnetic field is uniform and oriented along the z-axis; the white (resp. black) lines in panels A and C (resp. B) represent magnetic field lines. In all, the obstacle/chromosphere is located at the bottom, at z = 0, and t=0 corresponds to the moment when the stream hits the obstacle/chromosphere.

*Setup of the laboratory experiment and scalability to the astrophysical configuration*

In order to investigate the accretion dynamics of CTTSs in the laboratory, and as shown in Fig 1A, we create a collimated narrow (1 mm diameter) plasma stream by imposing a high-strength ($B_z$=20 T) external and uniform poloidal magnetic field onto an expanding plasma, ablated by a high- power laser (1 ns duration, $10^{13}$ W/cm$^2$) (*25,35*). The stream ($v_{stream}$=750 km/s, plasma electron density $2\times10^{18}$ $cm^{-3}$, plasma electron temperature 0.1 *MK* (*36*)) propagates parallel to the lines of the large-scale external magnetic field, as in the present picture of mass accretion in CTTSs, impacting onto an obstacle mimicking the high-density region of the star chromosphere (see *Methods*). The uniqueness of this laboratory experiment is not only in coupling laser-produced plasma with an externally applied, high-strength magnetic field, but also in being fully three-dimensional, and in allowing simultaneous optical and X-ray emission measurements (see Fig 1A). This setup differs in several notable ways from previous experiments focused on investigating radiative accretion shocks, where magnetic fields are totally absent (while the presence of the magnetic field is crucial in reproducing CTTSs observations in a realistic way), and in which the use of "shock-tubes" (*37,38*) affects the dynamics of the constrained plasma.

As detailed in the *Methods*, we have verified that between the two systems, in which the accretion shock is not radiation-dominated (*37*), dimensionless scaling parameters (*39*) are similar, and that the absence of relevant gravity in the laboratory does not preclude similarity since its effects will be significant only at late times. Regarding the β parameter (the ratio of the plasma pressure to the magnetic pressure), this varies in the laboratory from <1 for the incoming stream to >1 in the shocked region (see *Methods*). This is imposed by the accessible laboratory conditions (lower plasma density streams are more difficult to diagnose, and higher-strength magnetic field would lead to destructive setups). Such conditions are similar to what can be inferred in CTTSs when using stream densities as those typically inferred from X-ray observations ($10^{11}\ cm^{-3}$) and the lowest magnetic field values observed in CTTSs ($7\times10^{-4}\ T$). Such value of magnetic field strength is lower than expected as average in CTTSs ($7\times10^{-2}\ T$). However, we stress that there is a vast variety of physical conditions which are relevant to accretion streams in CTTSs. The physical case investigated here is hence representative of a class of streams with $\beta \geq 1$, and allows to unveil the evolution of the stream following its impact and the morphology of the post-shock plasma under the strong influence of the magnetic field. The verified scaling between the two systems leads to, in this case, the evolution of the laboratory plasma over ($10\ ns, 1\ mm$) replicating that in CTTSs over ($300\ s,\ 3\times10^4\ km$). Note that we also varied the laboratory magnetic field strength from 6 to 30 T and verified in all cases the formation of a distinct dense and ionized shell around the shocked core.

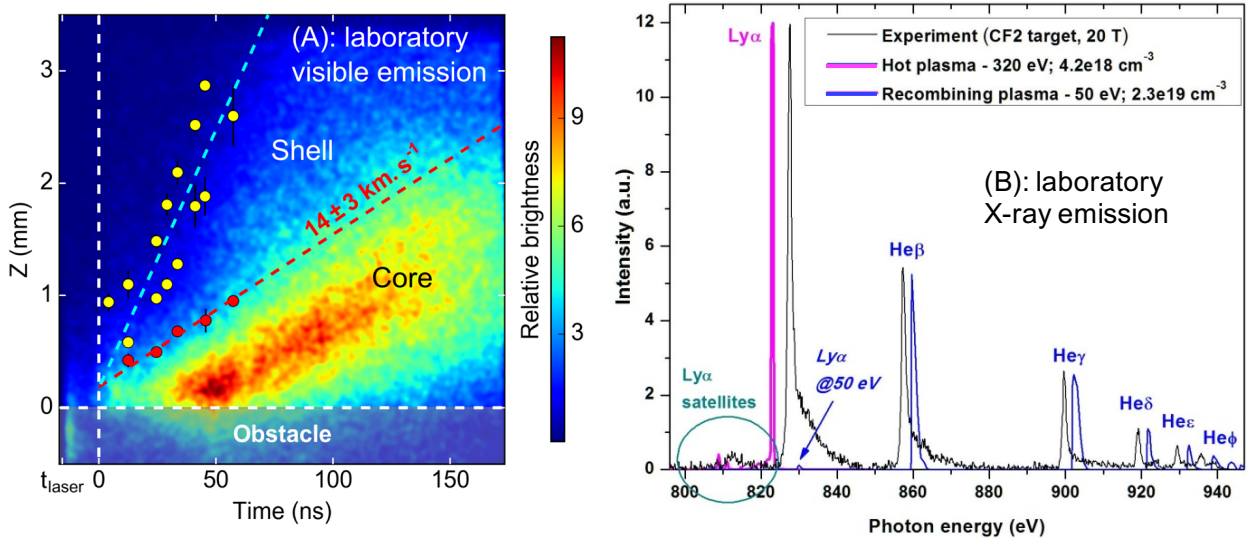

**Fig. 2: Visible and X-ray emissions produced simultaneously by the shocked core and shell plasmas and as recorded in the laboratory.** (A) Visible (time-and space-resolved, here the obstacle is a $CF_2$ target, while the stream is generated from a PVC $(C_2H_3Cl)_n$) laser-irradiated target) and (B) X-ray (integrated in time, and in space over 0 < z < 1 mm, i.e. near the obstacle, but spectrally resolved) emissions from the laboratory plasma. Note that here that, contrary to (A), the obstacle is a PVC $(C_2H_3Cl)_n$ target, while the stream is generated from a laser-irradiated $CF_2$

target. We however observe that the plasma density dynamics and characteristics (density, temperature) are the same whenever the laser target and obstacle targets are swapped. In (B), the configuration using a $CF_2$ stream-source target is used since the spectrometer records the spectrum corresponding to the Fluorine ions, and that most (95 %) of the plasma seen above the obstacle is composed of stream material, as precisely analysed by recording F-ions emission solely originating from stream or obstacle material. The spectrum shown in (B) uses the configuration of a $CF_2$ stream-source target since it leads to stronger emissions compared to when using the reverse configuration of a $CF_2$ obstacle target. Overlaid are the simulations of the emissions produced by two plasma components having the densities of the core and shell respectively, and temperatures of 0.58 MK (50 eV) and 3.7 MK (320 eV) respectively. The modeled spectra are offset along the photon energy scale for better visibility (note that the $Ly_\alpha$ line corresponds to the emission of H-like state F ions and that the He series to the emission of He-like state F ions).

*Results: a dense shell forms around the shocked core structure*

As shown in Fig 1B-D, what is observed in the laboratory and substantiated by the two MHD (in 2D and 3D) simulations (see *Methods*) is consistent with the following scenario: on impact, the stream, halted by the obstacle, induces the formation of an inward shock and of a reverse shock within the stream itself. The front of the reverse shock is localized by the density jump observed at the edge of the central core in Fig 1B-D. As we show in the *Supplementary Information*, we verify that in the laboratory the increase of the plasma electron density in the core post-shock region corresponds closely to what is expected from the Rankine-Hugoniot strong shock conditions (*27*). Simultaneously, the highly-conductive and heated post-shock core plasma is seen to expand laterally, its thermal pressure locally overcoming the magnetic field pressure and compressing the magnetic field lines (represented as the white lines in Fig 1D). Lateral expansion is halted by the increasing magnetic field, which induces the formation of a curved shock front that redirects the plasma flow toward the stream and forms an enveloping structure that we denote here as the "shell". In Fig 1B-D, we observe the ejected flow, the shell quickly overtaking (along z) the propagating reverse shock, as due to the longitudinal redirection of the ejected flow. We note that, despite the fact that the laboratory stream incident on the obstacle is observed to be steady over a longer duration than the time frames shown in Fig 1C, what limits its ability to represent a steady astrophysical column is the absence of gravity. This will affect (see *Methods*) the plasma dynamics at late times (e.g. the last frame of Fig 1C). However, at early times (i.e. the two first frames of Fig 1C), we can state that the plasma evolution observed in the laboratory, and hence the formation of the shell, is expected to be representative of steady accretion dynamics, the shell feature being observed to be steady in the simulations that include gravity.

As detailed in the *Methods* (Sec. 5), our simulations do not account for radiative transfer effects. This assumption can be considered valid only in the hot post-shock slab and in the corona (*20*). There the thermal conduction together with the radiative losses from optically thin plasma play a significant role in the energy budget. In particular, the intense radiative cooling at the base of the slab robs the post-shock plasma of pressure support, causing the material above the cooled layer to collapse. As a result, the shock position can vary in time (*24*). The thermal conduction acts as an additional cooling mechanism of the hot slab, draining energy from the shock-heated plasma to the chromosphere, and partially contrast the radiative cooling (*24*).

On the contrary, the cold and dense material of the stream and that of the chromosphere are most likely optically thick. As a result, the radiative transfer is expected to play a significant role in the energy budget whereas the thermal conduction should be negligible. The main effects are expected in the unshocked accretion column where the downfalling material can be radiatively heated up to temperatures of $10^5$ K (*20*). Also the optically thick material of the chromosphere and/or of the unshocked stream located along the line of sight is expected to partially absorb the X-ray emission arising from the hot post-shock slab. It is worth to mention that we do not consider the effects of radiative transfer on the dynamics and energetics of the system, but we account for the absorption in the synthesis of X-ray emission, as described in the *Methods* (Sec. 5). For this reason, our modeling is not entirely self-consistent. Nevertheless, we expect that the evolution of the post-shock plasma dominated by radiative cooling and thermal conduction is described accurately.

Both the shocked core plasma and the shell are simultaneously observed in the recorded laboratory plasma emissivities. The reverse shock front, and its temporal evolution, propagating up the stream at ~14±3 km/s, are clearly seen in the streaked visible emission (see *Methods*) of the laboratory plasma (Fig 2A): the reverse shock front identified in the density maps (the red points in Fig 2A) corresponds closely to the edge of the bright emitting, core post-shock region that expands toward the incoming stream. In the same emission map, we also clearly identify the shell, of reduced brightness, with its expansion front in the density maps (Fig 1C) identified by yellow points. Similarly, the X-ray laboratory emission originating near the obstacle surface (Fig 2B), as analysed by our non-steady model (*36*) and as detailed in the *Supplementary Information*, displays features characteristic of two distinct plasma components. Indeed, the appearance of remarkably intense He-series lines (from emitting He-like ionized F ions) is the witness of a plasma component having a low temperature, 0.6±0.1 MK, here at a density which corresponds well to the core density observed in Fig 1C. The simultaneous observation of a strong $Ly_\alpha$ line (from emitting H-like ionized F ions) attests the presence of a higher electron temperature plasma, i.e. at 3.7 MK as analysed using the atomic code FLYCHK (*40*). This synthetic radiation, as derived from the ratio between the $He_\beta$ and $Ly_\alpha$ line intensities, has volume and density consistent with the measured shell plasma (Fig 1C). Note that both shell and core temperatures derived this way are also well consistent with the laboratory simulation, as detailed in the *Supplementary Information*.

*Impact of the observed shell on X-rays emitted by the shocked core*

The dense and ionized shell, revealed here in the laboratory experiment, and also observed in the laboratory and astrophysical simulations, is seen in the latter case to modify the X-ray emission originating from the accretion region. From the X-ray emission computed from the simulation shown in Fig 1D, and comparing the results with and without local absorption effects (see *Methods*) taken into account (Fig 3), we indeed observe that the X-ray flux is significantly reduced when the local absorption is included. Our results thus reveal that taking into account the effect of the absorption by the dense and cold shell indeed participates to lowering the X-ray flux that can be observed originating from such stars, and thus influences the value of the mass accretion rate that can be inferred. The set of parameters accessible in the experiment, when scaled to the astrophysical case, corresponds to a situation where the obscuration effect (Fig 3C) is moderate. We can expect that for accretion streams characterized by high densities, a larger amount of optically thick material will surround the X-ray emitting slab, inducing heavier obscuration of emitted X-rays.

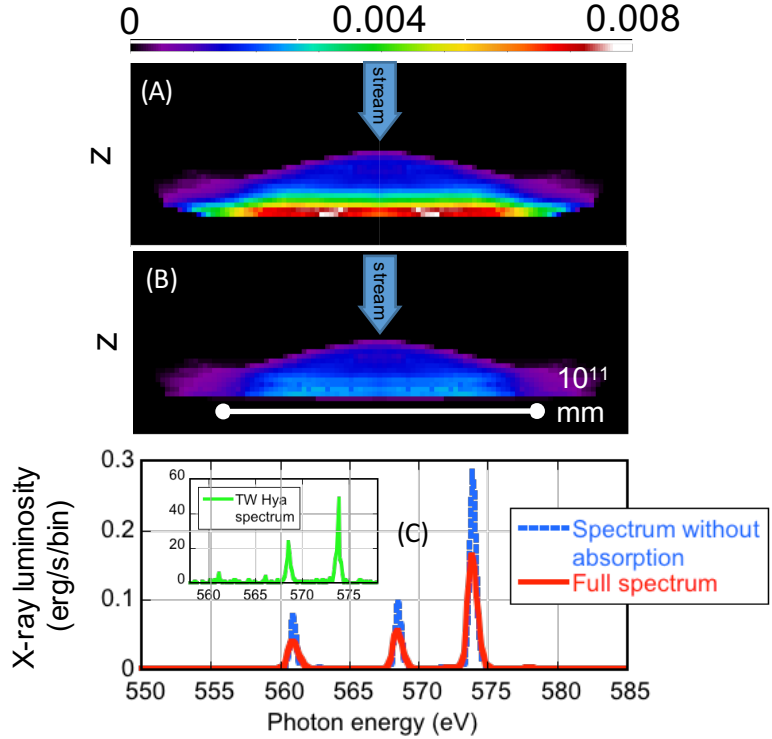

**Fig. 3: Simulation of reduced X-ray emissivity from a young star due to local absorption in the shell.** (A) X-ray emissivity maps (*22*) (the color bar is in erg/s per grid cell), as post-processed from the astrophysical simulation shown in Fig 1D, and looking along an axis perpendicular to the incident stream. (B) Same as (A), but taking into account the local absorption effect (see *Methods*). (C) The emitted spectrum, synthesized from the numerical model used for (A-B), in the energy range of the He-like OVII triplet and using the response function of the MEG grating of the Chandra satellite, with (red) and without (blue) the local absorption. Maps such as (A-B) are out of reach of observation capability, on the contrary to spectra which can be directly compared with astrophysical data such as the one shown in the inset of C, which displays the spectrum from the CTTS TW Hydrae, as observed by MEG/Chandra (*41*). The unit of the ordinates of the inset are in counts/bin.

*Conclusion and outlook*

Our results thus point to the crucial necessity to correctly account for plasma absorption in order to interpret and model accretion processes in young stars. Doing so allows for more accurate, with respect to alternate scenarios, modelling of observations (*21,21,22,23*), thus supporting the plausibility of the dynamics highlighted here and suggesting that indeed such conditions are present in CTTSs. The laboratory platform developed here also opens up the investigation of a number of other issues. For example, by changing the orientation of the stream with respect to the magnetic field, alternative channels of accretion can be explored (*42*).

**Acknowledgments:** We thank the LULI teams for technical support, the Dresden High Magnetic Field Laboratory at Helmholtz-Zentrum Dresden-Rossendorf for the development of the pulsed power generator, B. Albertazzi and M. Nakatsutsumi for their prior work in laying the groundwork for the experimental platform, and P. Loiseau for discussions. This work was supported by ANR Blanc Grant n° 12-BS09-025-01 SILAMPA (France) and by the Ministry of Education and Science of the Russian Federation under Contract No. 14.Z50.31.0007. This work was partly done within the LABEX Plas@Par project and supported by Grant No. 11-IDEX- 0004-02 from ANR (France). JIHT RAS and NRNU MEPhI members acknowledge the support of RFBR foundation in the frame of projects #14-29- 06099 and #15-32-21121 and the Competitiveness Program of NRNU MEPhI. O. Willi would like to acknowledge the DFG Programmes GRK 1203 and SFB/TR18. M.Borghesi and K. Naughton acknowledge funding from EPSRC (grants EP/J500094/1 and EP/P010059/1). AC, SO, RB, and CA acknowledge the support of PICS 6838 of CNRS. R. B. also acknowledges financial support from INAF under PRIN2013 Programme 'Disks, jets and the dawn of planets'. All data needed to evaluate the conclusions in the paper are present in the paper and/or the Supplementary Materials. Experimental data and simulations are respectively archived on servers at LULI, LERMA and INAF laboratories and can be consulted upon request. Part of the experimental system is covered by a patent (n°1000183285, 2013, INPI-France). PLUTO is developed at the Turin Astronomical Observatory in collaboration with the Department of Physics of Turin University.

J.F. and A.C. designed the project, J.F., D.P.H. and J.B. designed the experiment, J.B. and O.P. designed the pulsed high-magnetic field system, G.R., S.P., D.P.H., S.N.C., J.F., J.B., K.N., M.Bl., K.B., A.S., D.K., H.P. and R.R. performed the experiment, with M.Bo. and O.W. providing support. G.R., S.N.C., S.P., E.F., S.N.R., I.Y.S. and J.F. analysed the laboratory data, S.O., C.A., R. B., B.K., R.Ro. and A.C. designed and performed the simulations. J.F., G.R., S.N.C., S.O., C.A., R.B., S.P., A.C., B.K. wrote the article. All the authors provided comments on the various stages of the article.

Competing Interests: The authors declare that they have no competing interests.

## Materials and Methods:

### 1  Experimental Design: Experiment coupling lasers and magnetic field. Plasma generation.

The experiment was performed at the ELFIE Nd:glass laser facility of the LULI laboratory, at Ecole Polytechnique (France) (*43*), using its chirped laser pulse, delivering 40 J of energy within a 0.6 ns FWHM duration pulse at the wavelength $\lambda = 1057$ nm. This laser beam is focused on target over a 700 μm diameter flat-top spot, which gives an intensity of $I_{max}=1.6\times10^{13}$ W.cm$^{-2}$. The target is a flat solid from which, following laser irradiation, a hot plasma is created, expands into vacuum, and is funneled over 3 mm into a collimated stream (*25, 35, 44*) by the action of a large-scale, steady poloïdal external magnetic field (*34*) having a strength from 6 to 30 T, 20 T being the magnetic strength used for the results reported in the main text. The collimated plasma stream follows the magnetic field lines to hit a secondary solid target at a distance of 11.7 mm from the first target. When hitting this secondary obstacle target, the stream has a constant (over more than 100 ns) diameter of ~1.4 mm, and plasma electron density of ~$1.5\times10^{18}$ cm$^{-3}$ when using 20 T for the magnetic field and ~1.5 mm, and ~$1.2\times10^{18}$ cm$^{-3}$ for the same quantities when using 6 T for the magnetic field, as detailed in Tables S3 and S4 of the *Supplementary Information*. At 20 T, the measured plasma electron temperature is of ~10 eV, or 0.1 MK (*36,44*). We use two different materials, i.e. PolyVinyl Chloride (PVC, $(C_2H_3Cl)_n$) and Teflon ($CF_2$), for the two targets in order to be able to distinguish, through X-ray spectroscopy of the F-emitting ions (see below), the characteristics of the plasma coming from the impacting stream from the plasma ablated from the obstacle. These two materials are interchanged between shots between the primary and the secondary target, i.e. if the primary target is PVC, the secondary is Teflon, and vice-versa.

Both targets are situated at the center of a high-current Helmholtz coil system (*34*), constructed at the Laboratoire National des Champs Magnétiques Intenses (LNCMI) in Toulouse (France). The Helmholtz coil is inserted inside re-entrant tube enabling it to work at air, to allow heat dissipation and prevent any problems related to electric arcing (see also Fig S1 of the *Supplementary Information*). Two perpendicular apertures managed within the coil structure allow free access to the laser beams propagating in the vacuum of the target chamber and allow both targets to be inserted at the center of the coil, as well as the diagnostics to have a free line of sight to probe the plasma dynamics perpendicularly to the main axis of the stream expansion. The Helmholtz coils are coupled to a 32 kJ/16 kV capacitor bank that delivers 20 kA to the coils in order to generate up to a 20 T pulsed magnetic field in the center of the coils. The current discharge in the coils takes place over 204 µs (a half-period of a sinusoid), and the laser irradiation is synchronized with the peak of the discharge, i.e. when the magnetic field reaches its maximum strength. Hence, we can consider the magnetic field to be steady over several µs, which is much longer than the plasma dynamics investigated in the experiment (120 ns). The magnetic field is homogeneous in the longitudinal and radial directions over 40 mm (axially) and 20 mm (radially) respectively, which is much larger than the typical scales of the plasma dynamics (typically 12 mm axially and 5 mm radially).

## 2 Scalability between the laboratory and astrophysical plasmas

Scaling the laboratory flows to astrophysical flows relies on the two systems being described accurately enough by ideal MHD (*39*). The experiments rely on generating a relatively hot, conductive and inviscid, plasma so that the relevant dimensionless parameters are in the correct regime. Section 3 of the *Supplementary Information,* details the respective parameters of the two plasmas. For both laboratory and astrophysical flows respectively, in the initial conditions, i.e. in the incoming stream, the Peclet number, which is the ratio of heat convection to heat conduction, is 10 and $8\times10^8$ respectively, the Reynolds number, which is the ratio of the inertial force to the viscous one, is $4.6\times10^5$ and $2.6\times10^{11}$ respectively, and the magnetic Reynolds number, which is the ratio of the convection over ohmic dissipation, is 34 and $3.5\times10^9$ respectively. Hence, they are all ≫ 1 for the two systems. We also quantitatively verify that the two scaling quantities determined in Ref. (*39*), i.e. the Euler ( $v\sqrt{\frac{\rho}{p}}$, where v is the flow speed, ρ the density and p the pressure) number (40.8 in the laboratory case and 87 in the astrophysical case) and the Alfven ( $B/\sqrt{p}$ where B is the magnetic field) number ($1.1\times10^{-2}$ in the laboratory case and $1.2\times10^{-2}$ in the astrophysical case) are very similar between the two systems. This ensures that both MHD systems evolve in a similar way.

The scaling of the stream radius between the experiment and the CTTS simulation, which leads to 1 mm in the experiment corresponding to $3\times10^{10}$ mm ($3\times10^4$ km) in the astrophysical situation, is done as follows. We use as a basis the known quantities: stream radius (r), stream velocity (v) and stream density (ρ). Here we use the 20 T case shown in Fig 1C for the laboratory plasma, and the ones of the Pluto simulation shown in Fig 1D:

- r_lab=0.7 mm and r_astro=$0.5\times10^{11}$ mm,
- v_lab=100-1000 km/s (we measure only the 1000 km/s velocity of the detectable edge of the stream, its velocity decreasing with time as 1/t (*44*) and v_astro=500 km/s,
- ρ_lab=$9.7\times10^{-6}$ g/cm$^3$ and ρ_astro=$10^{-13}$ g/cm$^3$.

From these, we can retrieve the following scaling factors (*39*):

a= r_astro/ r_lab = $7.1\times10^{10}$

b= ρ_astro/ ρ_lab = $10^{-8}$

c= v_astro/ v_lab = 5-0.5

This allows to obtain the temporal scaling, which is: t_astro=(a/c)×t_lab. Hence, we can assess that, already for t>10 ns, when the stream velocity has dropped ≤100 km/s, a time 10 ns duration is equivalent to 143 s. This corresponds quite well to the respective temporal evolutions of the laboratory and astrophysically simulated plasmas observed in Fig 1.

We also note that we have a good correspondence with respect to the magnetic field scaling, which is: B_astro=$(c\sqrt{b})$ B_lab. At the initial time of the impact (i.e. using a stream velocity of 1000 km/s), this would yield: B_astro=$10^{-3}$ T for B_lab=20 T, which corresponds quite well with the value $7\times10^{-4}$ T used in the Pluto astrophysical simulation.

We note that the magnetization of the electrons and of the ions is in good correspondence between the laboratory and astrophysical situations (see Section 3 of the *Supplementary*

*Information*), where in the incoming stream, ions are not magnetized while the electrons are magnetized. For the rest of the dynamic, i.e. after the plasma has been shocked through the impact onto the surface, both species are magnetized. One difference between the astrophysical and laboratory dynamic is however the thermalization time between ions and electrons. In the stream, the ions have most of the ram energy, because of their mass compared to the electrons. They attain a higher gain in thermal energy through the shock by isotropisation of this ram energy. For the electrons, the temperature gain is almost negligible. The electrons can be heated only in a second time, by the ions, through collisions. Regarding the density and temperature conditions reached after the shock in both cases, one can see in the Section 3 of the *Supplementary Information* that the equilibration time in the laboratory shocked plasma is of the order of 30 ns, which is quite long compared with the time necessary for the particles to travel through the shocked region. However, in the astrophysical situation, this equilibration time is 2.5 s, which is actually negligible compared with the time necessary for the particles to travel through the shock. Consequently, in the laboratory plasma, the ion and electron temperature is decoupled during the dynamic of transit of both species, which are going through the core before to be ejected out of it, in order to form the shell (see main text). As a consequence, in the laboratory, the electron temperature is higher in the shell than in the core. Contrarily, in the astrophysical situation, the thermalization is fast enough for the electrons and ions to gain the same temperature quickly after the shock. The maximum temperature of the electrons is then already reached in the core, and the cooling by radiation leads to the temperature in the shell to be lower than in the core.

Finally, regarding the effect of the gravity, as it is absent in the laboratory experiment, we can state that in the laboratory case the effects of gravity on the post-shock flow (and on the plasma in general) are negligible over the duration of the experiments. In the astrophysical context however, gravitational forces are important and will tend to decelerate, over a time-scale $\tau_G \sim v_{bf}/g$ , the shell plasma that escapes from the shock regions and flows back along the accretion column; where $g = GM/R^2$ is the gravitational acceleration and $v_{bf}$ is the characteristic speed of the back-flow. The time-scale $\tau_G$ can be estimated by considering that the back-flowing plasma is driven by the thermal energy gained in the accretion shocks. Its characteristic speed will then be of the order of the sound speed in the post-shock, thus we can write $v_{bf} \sim c_s \sim \sqrt{k_B T \mu m_H} \sim \frac{1}{3} v_{ff}$, where for the last relation we have used the fact that for a strong shock the post-shock temperature is given by $k_B T \sim \frac{3}{32}\mu m_H v_{ff}^2$, where $v_{ff}$ is the free-fall speed of the accretion flow. Thus, upon impact of the accretion flow onto the stellar surface, gravity will become important for times $\gtrsim \tau_G \sim R/(3\ v_{ff})$. As an example, for values appropriate to the CTTS MP Mus, $R = 1.3$ R$_\odot$ and $M = 1.2$ M$_\odot$ (*6*), this time-scale is $\tau_G \sim 700$ s, which is indeed consistent with simulations done with gravity included and with what can be seen in Fig.1D of the main text.

## 3   Experimental diagnostics.

The plasma dynamics was diagnosed using mainly four complementary diagnostics: (1) an optical diagnostic using a short pulse laser coupled to an interferometer, allowing to measure in a snapshot the electron plasma density in the range from $5\times10^{16}$ to $10^{20}$ cm$^{-3}$ (*45*), (2) a monochromatic X-ray radiography diagnostic using a short-pulse laser-driven backlighter, and allowing to probe the plasma density in higher density regions than the optical probe (*46*), (3) a time-resolved, and one-dimensional space-resolved (along the main incoming jet and plasma expansion axis along the z-axis) measurement of the plasma self-emission in the visible range, i.e. exploiting Streaked Optical Pyrometry (SOP) (*47*), and (4) a X-ray spectrometer collecting the emission from the F-ions in the plasma in order to retrieve the plasma density and temperature (*36*). The first two diagnostics are 2D-space resolved and time-resolved, the third diagnostic is 1D-space resolved (along the z-axis) and time-resolved, the fourth diagnostic is 1D-space resolved (along the z-axis) and time-integrated.

The analysis techniques of the various diagnostics are detailed in the *Supplementary Information*.

## 4   Astrophysical Simulations.

The 2D astrophysical simulations are performed using the PLUTO code (*48*) which is a modular Godunov-type code for astrophysical plasmas. It provides a multiphysics, multialgorithm modular environment, which is detailed in the *Supplementary Information,* particularly oriented toward the treatment of astrophysical flows in the presence of discontinuities as in the case treated here.

The initial conditions of the simulations represent an accretion stream with constant plasma density and velocity, propagating through the stellar corona. The initial unperturbed stellar atmosphere is assumed to be magneto-static[1] and to consist of a hot (maximum temperature ≈ $10^6$ $K$) and tenuous ($n_H \approx 2\times10^8 cm^{-3}$) corona linked through a steep transition region to an isothermal chromosphere[2] that is uniformly at temperature $10^4$ $K$ and is $8.5\times10^8 cm$ thick. Initially the stream is in pressure equilibrium with the stellar corona and has a circular cross-section with a radius $r_{str} = 5\times10^9 cm$. We considered as reference the case of a stream with density and velocity compatible with those derived from the analysis of X-ray spectra of MP Mus (*6*), namely $n_{str0} = 10^{11} cm^{-3}$ and $u_{str0} = 500 km.s^{-3}$ at a height $z = 2.1\times10^{10} cm$ above the stellar surface; then we considered additional simulations slightly varying the value of stream density around the reference value. The stream temperature is determined by the pressure balance across the stream lateral boundary. The unperturbed stellar magnetic field is assumed to be uniform, aligned with the stream axis, and perpendicular to the stellar surface. Since the stream impact reproduced in the laboratory experiment has an evolution which resembles that of the intermediate run By-10 in (*24*), we considered magnetic field strengths that lead to a plasma β in the post-shock region similar to that of run By-10, namely ranging between 1 and 100. For the values of stream density explored, the magnetic field strength ranges between 7 x $10^{-4}$ and 5x$10^{-3}$ T. It is worth noting that the values of magnetic field strength used here are slightly lower than those expected in CTTSs (see main text). However, our primary goal was to reproduce the evolution of the stream impact and the morphology of the post-shock plasma as observed in the laboratory experiment. Both the evolution and the morphology are guided by the plasma-β value of the post-shock region which we reproduced using physical parameters as close as possible to the parameters characterizing stream impacts in CTTSs. A summary of all the simulations performed is given in Section 1 of the *Supplementary Information*.

## 5   Synthesis of the X-ray emission and comparison with astrophysical objects

For Fig 3 of the main paper, we synthesized the X-ray maps and the spectra from the D5e10-B07 model (see Section 1 of the *Supplementary Information*). We derived the maps and spectra both taking into account the local absorption of the surrounding medium along the line of sight (LoS) or neglecting these effects. This approach allows us to infer the role of the local

---

[1] We adapted the wind model of Ref. (*38*) to calculate the initial vertical profiles of mass density and temperature from the base of the transition region ($T = 10^4$ $K$) to the corona.
[2] Note that the radiative losses are set equal to zero in the chromosphere to keep it in equilibrium.

absorption on the detectability of the emission from the accretion shock.

The method to synthesize the maps and the spectra that can be directly compared with observations of accretion shocks consists of several steps of a tool properly developed to handle the high energy emission from shocks, as discussed in (*22*). From the bi-dimensional maps of the density, velocity, and temperature of the plasma simulated with the PLUTO code, we reconstruct the 3D maps by rotating the 2D slab around the symmetry z axis (reducing the original resolution of the numerical simulations). From the values of temperature and emission measure (EM) in each computational cell, assuming metal abundances of 0.5 of the solar values (in agreement with X-ray observations of CTTSs (*49*)), we synthesize the corresponding emission using the CHIANTI atomic database (*50*). We take into account the local absorption by computing the X-ray spectrum from each cell and by filtering it through the absorption column density along the LoS, i.e. we take into account the emission from each cell and the absorption of each cell in front along the line of sight (*22*). We use the absorption cross sections as a function of wavelength from (*51*) to compute the absorption, due mainly to cold material, as the soft X-ray opacity drops at high temperature (T > $10^6$ K (*52*)). We also subtract the emission from the coronal component and neglect the absorption due to the interstellar medium. We synthesize the X-ray maps and spectra emerging from the shock region by integrating the absorbed X-ray spectra from the cells in the whole domain.

It is worth noting that our astrophysical simulations do not include the effects of radiative transfer, so that they are not entirely self-consistent. However, the radiative transfer is expected to affect mostly the material of the pre-shock stream, developing a region of radiatively heated gas (a precursor) in the unshocked accretion column (*20*). As such, this heating mechanism does not affect significantly the dynamics of the post-shock plasma (*20*).

# 6    Simulation of the laboratory experiment using the GORGON code

The laboratory simulations of the accretion experiments are performed using the 3D resistive MHD code GORGON (*53,54*), a highly parallel 3D MHD code. The model describes a bi-temperature, single-fluid resistive plasma in an optically thin regime, with isotropic Braginskii-like transport coefficients. We modeled the entire experiment (as detailed in the *Supplementary Information*, Section 7), starting from the initial expansion of the laser-produced plasma from the first laser-irradiated target, to the impact and accretion of the collimated stream on the obstacle.

The initial precursor plasma evolution, as generated by the laser irradiation, (up to 1 *ns*) is modelled in axisymmetric, cylindrical geometry with the two-dimensional, three-temperatures, Lagrangian, radiation hydrodynamic code DUED (*55*). The laser parameters are taken to be similar to the experimental ones. The magnetic field is initially perpendicular to both laser and obstacle targets and has a magnitude of 20 *T*. We consider "outflow" boundary conditions for the flow, and for the magnetic field a continuous perpendicular component.

**List of Supplementary materials:**

Supplementary Information

Table S1 – S4

Fig S1 – S14

Movie S1 – S2

References 56 to 80 are only used in the Supplementary Information

**References:**

[1] A. Caratti o Garatti, et al., Disk-mediated accretion burst in a high-mass young stellar object. *Nat. Phys*. http://dx.doi.org/10.1038/ nphys3942 (2016).

[2] S. Scaringi, *et al*., Accretion-induced variability links young stellar objects, white dwarfs, and black holes. *Science Advances* **1**, e1500686 (2015).

[3] J. Muzerolle, *et al*., Unveiling the Inner Disk Structure of T Tauri Stars *Astrophys. J. Lett*. **597**, L149-L152 (2003).

[4] M. Camenzind, *Magnetized Disk-Winds and the Origin of Bipolar Outflows*, in *Accretion and Winds. Reviews in Modern Astronomy*, G. Klare, Eds. (Springer, Berlin, Heidelberg, 1990), vol. 3, pp. 234-265.

[5] G. G. Sacco *et al*., On the observability of T Tauri accretion shocks in the X-ray band. *Astron. Astrophys.* **522**, A55 (2010).

[6] G. Argiroffi, A. Maggio, G. Peres, X-ray emission from MP Muscae: an old classical T Tauri star. *Astron. Astrophys.* **465**, L5 (2007).

[7] C. Argiroffi *et al.*, The close T Tauri binary system V4046 Sgr: rotationally modulated x-ray emission from accretion shocks. *Astrophys. J*. **752**, 100 (2012).

[8] N. S. Brickhouse *et al*., A deep chandra x-ray spectrum of the accreting young star TW Hydrae. *Astrophys. J*. **710**, 1835 (2010).

[9] Joel H. Kastner *et al*., Evidence for accretion: high-resolution x-ray spectroscopy of the classical T Tauri star TW Hydrae. *Astrophys. J*. **567**, 434 (2002).

[10] B. Stelzer, J. H. M. M. Schmitt, X-ray emission from a metal depleted accretion shock onto the classical T Tauri star TW Hya. *Astron. Astrophys*. **418**, 687 (2004).

[11] R. L. Curran *et al.*, Multiwavelength diagnostics of accretion in an x-ray selected sample of CTTSs. *Astron. Astrophys.* **526**, A104 (2011).

[12] G.J. Herczeg & L.A. Hillenbrand, UV Excess Measures of Accretion onto Young Very Low Mass Stars and Brown Dwarfs. *Astrophys. J.* **681**, 594 (2008).

[13] E. Gullbring *et al.*, Disk Accretion Rates for T Tauri Stars. *Astrophys. J.* **492**, 323 (1998).

[14] J. H. M. M. Schmitt *et al.*, X-rays from accretion shocks in T Tauri stars: The case of BP Tau. *Astron. Astrophys.* **432**, L35 (2005).

[15] H. M. Günther, J. H. M. M. Schmitt, J. Robrade and C. Liefke, X-ray emission from classical T Tauri stars: accretion shocks and coronae? *Astron. Astrophys.* **466**, 1111 (2007).

[16] C. Argiroffi *et al.*, X-ray optical depth diagnostics of T Tauri accretion shocks. *Astron. Astrophys.* **507**, 939 (2009).

[17] T. Matsakos *et al.*, YSO accretion shocks: magnetic, chromospheric or stochastic flow effects can suppress fluctuations of X-ray emission. *Astron. Astrophys.* **557**, A69 (2013).

[18] S. Colombo, S. Orlando, G. Peres, C. Argiroffi, F. Reale, Impacts of fragmented accretion streams onto Classical T Tauri Stars: UV and X-ray emission lines. *Astron. Astrophys.* **594**, A93 (2016).

[19] G. G. Sacco *et al.*, X-ray emission from dense plasma in classical T Tauri stars: hydrodynamic modeling of the accretion shock. *Astron. Astrophys.* **491**, L55 (2008).

[20] G. Costa, S. Orlando, G. Peres, C. Argiroffi, R. Bonito, Hydrodynamic Modeling of Accretion Impacts in Classical T Tauri Stars: Radiative Heating of the Pre-shock Plasma. *Astron. Astrophys.* **597**, A1 (2017).

[21] F. Reale *et al.*, Bright hot impacts by erupted fragments falling back on the sun: a template for stellar accretion. *Science* **341**, 251 (2013).

[22] R. Bonito *et al.*, Magnetohydrodynamic modeling of the accretion shocks in classical T Tauri stars: the role of local absorption in the x-ray emission. *Astrophys. J.* **795**, L34 (2014).

[23] S. Orlando *et al.*, Radiative accretion shocks along nonuniform stellar magnetic fields in classical T Tauri stars. *Astron. Astrophys.* **559**, A127 (2013)

[24] S. Orlando *et al.*, X-ray emitting MHD accretion shocks in classical T Tauri stars: case for moderate to high plasma-$\beta$ values. *Astron. Astrophys.* **510**, A71 (2010).

[25] B. Albertazzi *et al.*, Laboratory formation of a scaled protostellar jet by coaligned poloidal magnetic field. *Science* **346**, 325-328 (2014).

[26] B. A. Remington, David Arnett, R. P. Drake, Hideaki Takabe, Modeling Astrophysical Phenomena in the Laboratory with Intense Lasers. *Science* **284**, 1488 (1999).

[27] R. P. Drake, Eds. *High Energy Density Physics*, (Springer, Berlin, Heidelberg, 2006).

[28] H. Daido *et al.*, Ultrahigh Pulsed Magnetic Field Produced by a CO2 Laser. *Jpn. J. Appl. Phys.* **26**, 1290 (1987).

[29] C. Courtois, A. D. Ash, D. M. Chambers, R. A. D. Grundy, and N. C. Woolsey, Creation of a uniform high magnetic-field strength environment for laser-driven experiments. *Journal of Applied Physics* **98**, 054913 (2005).

[30] B.B. Pollock *et al.*, High magnetic field generation for laser-plasma experiments. *Rev. Sci. Instrum.* **77**, 114703 (2006).

[31] D.H. Froula *et al.*, Quenching of the Nonlocal Electron Heat Transport by Large External Magnetic Fields in a Laser-Produced Plasma Measured with Imaging Thomson Scattering. Phys. Rev. Lett. **98**, 135001 (2007).

[32] O. V. Gotchev, *et al.*, Laser-Driven Magnetic-Flux Compression in High-Energy-Density Plasmas. Phys. Rev. Lett. **103**, 215004 (2009).

[33] G. Fiksel *et al.*, Magnetic Reconnection between Colliding Magnetized Laser-Produced Plasma Plumes. *Phys. Rev. Lett.* **113**, 105003 (2014).

[34] B. Albertazzi *et al.*, Production of large volume, strongly magnetized laser-produced plasmas by use of pulsed external magnetic fields. *Rev. Sci. Instrum.* **84**, 043505 (2013).

[35] A. Ciardi et al., Astrophysics of magnetically collimated jets generated from laser-produced plasmas. *Phys. Rev. Lett.* **110**, 025002 (2013).

[36] S. N. Ryazantsev *et al.*, Diagnostics of laser-produced plasmas based on the analysis of intensity ratios of He-like ions x-ray emission. *Phys. Plasmas* **23**, 123301 (2016).

[37] R.P. Drake *et al.*, Radiative effects in radiative shocks in shock tubes. *High Energy Density Phys.* **7**, 130-140 (2011).

[38] J. E. Cross *et al.*, Laboratory analogue of a supersonic accretion column in a binary star system. *Nat. Commun.* **7**, 11899 (2016).

[39] D. D. Ryutov, R. P. Drake, B. A. Remington, Criteria for scaled laboratory simulations of astrophysical MHD phenomena. *Astrophys. J. Suppl. Ser.* **127**, 465–468 (2000).

[40] H.-K. Chung, M. H. Chen, W. L. Morgan, Y. Ralchenko, R. W. Lee, FLYCHK: Generalized population kinetics and spectral model for rapid spectroscopic analysis for all elements. *High Energy Density Phys.* **1**, 3-12 (2005).

[41] N. S. Brickhouse, S. R. Cranmer, A. K. Dupree, G. J. M. Luna, S. Wolk, A Deep Chandra X-Ray Spectrum of the Accreting Young Star TW Hydrae. *Astrophys. J.* **710**, 1835 (2010).

[42] S. Casassus, et al., Flows of gas through a protoplanetary gap. *Nature* **493**, 191 (2013).

[43] J. P. Zou *et al.*, Recent progress on LULI high power laser facilities. *J. Phys.: Conf. Ser.* **112,** 032021 (2008).

[44] D. P. Higginson *et al.*, Detailed Characterization of Laser-Produced Astrophysically-Relevant Jets Formed via a Poloidal Magnetic Nozzle, *High Energy Density Phys.* **23**, 48-59 (2017).

[45] D. W. Sweeney, D. T. Attwood, and L. W. Coleman. Interferometric probing of laser produced plasmas. *Appl. Opt.* **15**, 1126-1128 (1976).

[46] H.-S. Park *et al.*, High-energy Kα radiography using high-intensity, short-pulse lasers. *Phys. Plasmas* **13**, 056309 (2006).

[47] N. C. Woolsey *et al.*, Laboratory plasma astrophysics simulation experiments using lasers. *J. Phys.: Conf. Ser.* **112**, 042009 (2008).

[48] A. Mignone, *et al.*, PLUTO: A Numerical Code for Computational Astrophysics. *Astrophys. J. Suppl. Ser.* **170**, 228 (2007).

[49] A. Telleschi, M. Güdel, K. R. Briggs, M. Audard, L. Scelsi, Special feature high-resolution x-ray spectroscopy of T Tauri stars in the Taurus-Auriga complex. *Astron. Astrophys.* **468**, 443-462 (2007).

[50] E. Landi, G. Del Zanna, P. R. Young, K. P. Dere, H. E.Mason, Chianti—an Atomic Database for Emission Lines. Xii. Version 7 of the Database. *Astrophys. J.* **744**, 99 (2012).

[51] M. Balucinska-Church, D. McCammon, Photoelectric absorption cross sections with variable abundances. *Astron. Astrophys* **400**, 699 (1992).

[52] J. H. Krolik, T. R. Kallman, Soft x-ray opacity in hot and photoionized gases. *Astrophys. J.* **286**, A69 (1984).

[53] A. Ciardi *et al.*, The evolution of magnetic tower jets in the laboratory. *Phys. Plasmas* **14**, 056501 (2007).

[54] J. P. Chittenden, S. V. Lebedev, C.A. Jennings S. N. Bland and A Ciardi, X-ray generation mechanisms in three-dimensional simulations of wire array Z-pinches. *Plasma Physics and Controlled Fusion* **46**, B457 (2004).

[55] S. Atzeni *et al.*, Fluid and kinetic simulation of inertial confinement fusion plasmas. *Comput. Phys. Commun*. **169**, 153 (2005).

[56] E. Anders, N. Grevesse, Abundances of the elements: meteoritic and solar. *Geochim. Cosmochim. Acta*, **53**, 197 (1989).

[57] V. Kashyap, J. J. Drake, PINTofALE: Package for the interactive analysis of line emission. Bull. Astr. Soc. India **28**, 475-476 (2000).

[58] R. K. Smith, N. S. Brickhouse, D. A. Liedahl, J. C. Raymond, Collisional plasma models with APEC/APED: emission-line diagnostics of Hydrogen-like and Helium-like ions. *Astrophys. J.* **556**, L91-L95 (2001).

[59] L. Spitzer, Eds., *Physics of Fully Ionized Gases* (Interscience, New York, 1956).

[60] W. W. Dalton, S. A. Balbus, A Flux-limited Treatment for the Conductive Evaporation of Spherical Interstellar Gas Clouds. *Astrophys. J.* **404**, 625-635 (1993).

[61] S. Orlando, F. Bocchino, F. Reale, G. Peres, P. Pagano, The importance of magnetic-field-oriented thermal conduction in the interaction of SNR shocks with interstellar clouds. *Astrophys. J.* **678**, 274 (2008).

[62] L. L. Cowie, C. F. McKee, The evaporation of spherical clouds in a hot gas. I - Classical and saturated mass loss rates. *Astrophys. J.* **211**, 135-146 (1977).

[63] J. L. Giuliani Jr., On the dynamics in evaporating cloud envelopes. *Astrophys. J.* **277**, 605G (1984).

[64] K. J. Borkowski, J. M. Shull, & C. F. McKee, Two-temperature radiative shocks with electron thermal conduction. *Astrophys. J.* **336**, 979B (1989).

[65] Yu. A. Fadeyev, H. Le Coroller, D. Gillet, The structure of radiative shock waves. IV. Effects of electron thermal conduction. *Astron. Astrophys.* **392**, 735F (2002).

[66] T. Vinci, A. Flacco, Neutrino (2014), URL https://github.com/aflux/neutrino.

[67] T. Pisarczyk, R. Arendzikowski, P. Parys, Z. Patron, Polaro-interferometer with automatic images processing for laser plasma diagnostic. *Laser Part. Beams* **12**, 549 (1994).

[68] Ya. B. Zel'Dovich and Yu. P. Raizer, *Physics of shock waves and high- temperature hydrodynamic phenomena*, (New York Academic Press, 1967).

[69] S.N. Chen *et al.*, Density and temperature characterization of long-scale length, near-critical density controlled plasma produced from ultra-low density plastic foam. *Sc. Rep.* **6**, 21495 (2016).

[70] A. L. Meadowcroft, C. D. Bentley, E. N. Stott, Evaluation of the sensitivity and fading characteristics of an image plate system for x-ray diagnostics. *Rev. Sci. Instrum.* **79**, 113102 (2006).

[71] Stefano Atzeni, and Jürgen Meyer-ter-Vehn, *The Physics of Inertial Fusion: Beam Plasma Interaction, Hydrodynamics, Hot Dense Matter* (Oxford University Press, 2009).

[72] A.Ya Faenov et al., High-performance x-ray spectroscopic devices for plasma microsources investigations. Phys. Scr. **50**, 333 (1994).

[73] J. J. MacFarlane, I. E. Golovkin, P. Wang, P. R. Woodruff, N. A. Pereira, SPECT3D – A multi-dimensional collisional-radiative code for generating diagnostic signatures based on hydrodynamics and PIC simulation output, *High Energy Density Phys.* **3,** 181 (2007).

[74] D. D. Ryutov, et al., Similarity Criteria for the Laboratory Simulation of Supernova Hydrodynamics. *Astrophysical Journal* **518**, 821 (1999).

[75] M. Dorf, https://arxiv.org/abs/1408.3677

[76] Ralph S. Sutherland, and M. A. Dopita, Cooling functions for low-density astrophysical plasmas. *Astrophysical Journal Supplement Series* **88**, 253-327 (1993).

[77] J. D. Huba, NRL Plasma Formulary. Washington, DC: Naval Research Laboratory, 1998.

[78] L. Mejnertsen, J. P. Eastwood, J. P. Chittenden, and A. Masters, Global MHD simulations of Neptune's magnetosphere. *J. Geophys. Res. Space Physics* **121**, 7497–7513 (2016).

[79] D. Salzmann, Atomic physics in hot plasmas (Oxford University Press, 1998).

[80] E.M. Epperlein, & M.G. Haines, Plasma transport coefficients in a magnetic field by direct numerical solution of the Fokker–Planck equation. *Physics of Fluids* **29**, 1029 (1986).

[81] S.I. Braginskii, Transport Processes in Plasma. *Reviews of Plasma Physics* **1**, 205 (1965).